\magnification=\magstep1

\hsize=150truemm
%
\parindent=0pt
\parskip=20pt
\def\hook{\hbox{\vrule height0pt width4pt depth0.5pt
\vrule height7pt width0.5pt depth0.5pt \vrule height0pt width2pt
depth0pt} }
\pageno=-1
\centerline{\bf Dirac Symmetry Operators }
\vskip 0.2true cm
\centerline{\bf From }
\vskip 0.2true cm
\centerline{\bf Conformal Killing-Yano Tensors.}
\vskip 1true cm
\centerline{ {\bf I. M. Benn}\footnote{${}^1$}
{email:mmimb@cc.newcastle.edu.au}}
\centerline{{\bf Philip Charlton}\footnote{${}^2$}
{email:philipc@maths.newcastle.edu.au}}
\vskip 0.5true cm
$$\vbox{\halign{\hfil#\cr        
Mathematics Department,\cr
Newcastle University,\cr
NSW 2308,\cr
Australia.\cr}}$$

\vskip 1true cm
\noindent{\bf Abstract.}  We show how, for all dimensions and signatures,
a symmetry operator for the massless Dirac equation can be constructed
from a conformal Killing-Yano tensor of arbitrary degree. 

\vfill\eject
\pageno=1

A symmetry operator of an equation maps any solution to another. 
On a (pseudo) Riemannian manifold isometries give rise to symmetry 
operators; the Lie derivative with respect to a Killing vector being a 
symmetry  operator  for  differential equations constructed from the
metric and the connection.  For those equations having a
conformal covariance the 
conformal Killing vectors also give symmetry operators.
Killing's equation for isometry generators has been generalised to  
tensor equations. One generalisation is to a totally symmetric Killing
tensor,
whilst Yano's generalisation of  Killing's equation applies to
a totally antisymmetric tensor, or differential form [1]. 
The conformal generalisation of Yano's Killing tensor equation
was given by Tachibana [2], for the case of a 2-form, and more generally
by Kashiwada [3], [4].  

Of direct physical interest is the case of a four-dimensional 
Lorentzian-signatured space-time, and it is in this situation that 
most of the 
applications of conformal  Killing-Yano tensors have  taken place. 
(A review of Killing-Yano tensors in relativity is given in [5].)
In particular, Kamran and McLenaghan [6] have shown that a symmetry
operator for the massless Dirac equation can be made from a conformal
Killing-Yano tensor. In the more restrictive case of a Killing-Yano
tensor this operator is a symmetry operator of the massive Dirac equation
as was previously found	by Carter and McLenaghan [7]. 

There are  obviously restrictions on the degree of  conformal Killing-Yano
tensors	possible in four dimensions.
The 0-form and 4-form cases are trivial, with conformal
Killing-Yano tensors being constant (parallel) in this case. 
The conformal Killing-Yano (CKY) equation for a	1-form  is just the 
conformal Killing equation. Since the CKY equation is invariant under 
Hodge duality  any  CKY 3-form  is just the dual of a conformal 
Killing vector. So in four dimensions
the only non-trivial generalisation of a conformal Killing vector 
is to a 2-form.  Because 2-forms are `middle-forms' in four dimensions,
they have properties not shared by a general CKY $p$-form. In particular
the null self-dual middle-forms correspond to maximal isotropic planes.
In the Lorentzian case these (complex) null foliations have an associated
(real) null shear-free congruence. This distribution of isotropic planes
can also be put into correspondence with a spinor field, and more generally
a non-null CKY 2-form corresponds to a pair of spinor fields. A CKY
2-form corresponds to a `tensor-spinor'  known as a `Killing spinor'. 
It was using this spinorial version of the equation that Torres del Castillo
[8] independently found the symmetry operator of the massless Dirac equation
that had been found by Kamran and McLenaghan [6]. CKY tensors can only
exist on special manifolds. In the four-dimensional Lorentzian case 
integrability conditions for the CKY equation can be written in terms of
the conformal tensor, and this restricts the Petrov type of the space-time.
One cannot interpret the integrability conditions of the CKY equation
in this way in the more  general setting. Thus it is not immediately
clear that the symmetry operator for the massless Dirac equation made
from a CKY 2-form in four dimensions will generalise to the case of
a $p$-form in $n$ dimensions.

In this paper we will consider a CKY tensor of arbitrary degree on 
a pseudo-Riemannian manifold of arbitrary dimension and signature. 
We will show how 
such a tensor gives a symmetry operator for the massless Dirac equation.
This symmetry operator generalises that given in the four-dimensional
case by Kamran and McLenaghan [6]. 
We follow the conventions and notation used in [9], and more generally 
 [10].

The CKY equation can be written most compactly in terms of exterior
operations. A $p$-form $\omega$ is a conformal Killing-Yano (CKY) tensor 
in $n$ dimensions
if its covariant derivative is related to the exterior derivative $d$ and
coderivative $d^*$ by  the equation
$$\nabla_X\omega = {1\over p+1}X\hook d\omega - {1\over n-p+1}X^\flat
\wedge d^*\omega \,. \eqno(1)$$
Here $X^\flat$ denotes the 1-form  obtained by `lowering' the components
of the vector field $X$ with the metric tensor $g$. 
(For the case of a 2-form in four dimensions we explicitly showed in [9]
the correspondence between this exterior equation and that originally
given by Tachibana.) 
Since the coderivative is related to the exterior derivative and the Hodge 
dual  by
$$d^*\omega = (-1)^p*^{-1}d*\omega = (-1)^{n-p-1}*d*^{-1}\omega\,,\eqno(2)$$
it follows that the CKY equation (1) is invariant under Hodge duality; that 
is, if $\omega$ satisfies the equation then so does $*\omega$. 
The equation is also covariant under conformal rescalings.
If $\omega$ is a CKY tensor on a manifold with metric tensor $g$, then
$\hat\omega$ is a CKY tensor with the conformally scaled metric $\hat g$
where the conformal weights are related by
$$\eqalign{
\hat g &= e^{2\lambda} g \cr
\hat\omega &=e^{(p+1)\lambda}\omega \,.\cr}\eqno(3)$$
In the case in which  there is a  conformal Killing vector $k$ 
then the Lie derivative ${\cal L}_k$ 
gives a symmetry operator of the CKY equation.
If $\omega$ is a CKY $p$-form then so is 
$$\eqalign{ 
{\cal L}_k\omega &-(p+1)\lambda\omega  \cr
\hbox{where}\quad {\cal L}_kg &=2\lambda g\,.\cr}\eqno(4)$$

By differentiating (1) we obtain an expression for the curvature operator
$R(X,Y)$ acting on $\omega$. If we then note that the exterior 
derivative and 
coderivative can be expressed in terms of the
(pseudo) Riemannian connection as
$$\eqalignno{
d &= e^a\wedge\nabla_{X_a} \cr
d^* &= -X^a\hook\nabla_{X_a}\,,\cr}$$
then we obtain the integrability condition
$${p\over p+1}d^*d\omega +{(n-p)\over(n-p+1)}dd^*\omega = e^b\wedge X^a\hook
R(X_a,X_b)\omega \,. \eqno(5)$$

Spinor fields carry irreducible representations of the 
Clifford algebra, elements of which can be identified with differential
forms. We denote this Clifford action by juxtaposing symbols.  
Thus for example we write the Dirac operator $D$ as
$$D = e^a\nabla_{X_a}\,.$$
The juxtaposing of two differential forms will denote their Clifford
product. 
Out of a CKY tensor $\omega$ we construct an operator on spinor fields:
$$L_\omega =e^a\omega\nabla_{X_a} + {p\over p+1}d\omega 
-{(n-p)\over(n-p+1)}d^*\omega  \,. \eqno(6)$$
To see that we have a symmetry operator we compose $L_\omega$ with the
Dirac operator which we bring through from left to right.
The integrability condition (5)	can be used to write the terms 
containing second derivatives of $\omega$ in terms of the 
curvature operator, whilst the 
second-order derivative operator can be written in terms
of the square of the Dirac operator, related to the spinor Laplacian by
$$D^2 =\nabla^2 +{1\over2}e^{ab}R(X_a,X_b)\,,$$
where we write $e^{ab}$ as a shorthand for $e^a\wedge e^b$. Thus we obtain
$$\eqalignno{
DL_\omega &= \omega D^2 +{(-1)^p\over p+1}d\omega D 
+{(-1)^p\over n-p+1}d^*\omega D  \cr
&\quad +{1\over2}[e^{ab},\omega]R(X_a,X_b)
+ e^b\wedge X^a\hook R(X_b,X_a)\omega\,. &(7)\cr
}$$
Now we may expand the curvature operator terms as follows:
$$\eqalignno{
[e^{ab},\omega]R(X_a,X_b)\psi &= e^{ab}\omega R(X_a,X_b)\psi -
\omega e^{ab}R(X_a,X_b)\psi \cr
&=e^{ab}R(X_a,X_b)(\omega\psi) -e^{ab}R(X_a,X_b)\omega\psi 
-\omega e^{ab}R(X_a,X_b)\psi\,.\cr}$$
The spinor curvature operator is given in terms of the curvature 2-forms
by 
$$\eqalignno{
R(X_a,X_b)\psi &={1\over2}R_{ab}\psi  \cr
\noalign{\hbox{and so}}
e^{ab}R(X_a,X_b)\psi &= -{1\over2}{\cal R}\psi \cr}$$
where $\cal R$ is the curvature scalar. Thus
$$\eqalignno{
[e^{ab},\omega]R(X_a,X_b)\psi &=-e^{ab}R(X_a,X_b)\omega\psi \cr
&= 2e^a\wedge X^b\hook R(X_a,X_b)\omega \psi \cr}$$
and so the curvature terms in (7) cancel. Thus letting both sides of (7)
operate on a solution of the massless Dirac equation shows that $L_\omega$
is a symmetry operator. 

As we previously noted, the CKY equation is duality invariant. To relate
the symmetry
operators obtained from $\omega$ and its dual $*\omega$
we use the relation between the Hodge dual and Clifford multiplication
by the volume $n$-form $z$,
$$*\omega = (-1)^{[p/2]}\omega z $$
where the square brackets denote the integer part [10]. If we then use (2)
we see that
$$L_{*\omega} = (-1)^{[p/2]}L_\omega z \,. \eqno(8)$$
Solutions to the massless Dirac equation decompose into semi-spinors
that are eigenvectors under multiplication by $z$. Thus $L_\omega$ and
$L_{*\omega}$ only differ by a scaling when acting on these semi-spinors.

Obviously any operator proportional to the Dirac operator is a trivial symmetry
operator.  By adding such a term to  $L_\omega$   we obtain a symmetry operator
whose commutator with the Dirac operator vanishes when $\omega$ satisfies the
stronger requirement of being a Killing-Yano tensor. If we define $K_\omega$ by
$$K_\omega = L_\omega - (-1)^p\omega D \eqno(9)$$ then we have $$\eqalignno{
\{D, K_\omega\} &={2(-1)^p\over (n-p+1)}d^*\omega D &(10)\cr
\noalign{\hbox{where $\{D,K_\omega\}$ is the graded commutator,}}
\{D,K_\omega\} &= DK_\omega+(-1)^pK_\omega D\,. \cr}$$ The right-hand side of
(10) vanishes when  $d^*\omega=0$, which is just the condition that $\omega$ be 
a Killing-Yano tensor.  In the case in which $\omega$ is a 1-form it is the
dual of a conformal Killing vector, $\omega=k^\flat$. Then if 
$$\eqalignno{
{\cal L}_kg &= 2\lambda g\,,\cr d^*k^\flat &= -n\lambda \cr \noalign{\hbox{and}}
K_{k^\flat} &= 2\left({\cal L}_k +{(n-1)\over2}\lambda\right) \cr
\noalign{\hbox{where the Lie derivative on spinors is defined by [10]}} {\cal
L}_k &=\nabla_k +{1\over4}dk^\flat \,. \cr}$$

In even dimensions the volume form $z$ anti-commutes with the Dirac operator.
Thus when $\omega$ is odd (even) $K_\omega$ preserves (interchanges) the 
semi-spinor spaces; that is, the eigenspaces of $z$. 
When $\omega$ is even then the anti-commutator in (10) can be changed to
a commutator by replacing $K_\omega$ with $K_\omega z$. Thus in even 
dimensions any Killing-Yano tensor gives an operator that commutes with
the Dirac operator, and hence is a symmetry operator of the massive Dirac
equation.

In odd dimensions  the volume form commutes with everything.
Thus when $\omega$ is even we cannot change the anti-commutator in (10)
to a commutator as we did in the even-dimensional case. 
Since in odd dimensions the dual of an even co-closed form is an odd 
closed one, one might try changing $K_\omega$
in such a way that the coderivative in (10) is replaced with the exterior
derivative. That is, we may define $K'_\omega$ by
$$K'_\omega = L_\omega +(-1)^p\omega D \eqno(11)$$
such that we have
$$DK'_\omega -(-1)^pK'_\omega D = {2(-1)^p\over p+1}d\omega D\,.\eqno(12)$$
However, in modifying $K_\omega$ such that the coderivative in (10) is
replaced with the derivative, the sign in the graded commutator	is also
changed. Thus for $K'_\omega$ to commute with the Dirac operator $\omega$
must be even and closed: that is, its dual is odd  and coclosed. 
So $K'_\omega$ offers nothing new. 
To have a symmetry operator of the massive Dirac equation we need
an odd coclosed CKY tensor, a Killing-Yano tensor, or an even closed one.

In the four-dimensional Lorentzian case the symmetry operator $K_\omega$ 
is  that given by Kamran and McLenaghan. We wrote it in this form in [9]
where we showed the relation to `Debye potentials' for the Dirac equation.
The operator $K_\omega$ is said to be $R$-commuting (or anti-commuting)
as the (graded) commutator with the Dirac operator is of the form
$\{K_{\omega},D\}= RD$. Kamran and McLenaghan [6] 
obtained (in four dimensions)
the most general first-order $R$-commuting  operator for the Dirac
operator. They showed that the non-trivial terms were made out of a
conformal Killing vector and a CKY tensor; that is, CKY tensors of all
possible degrees. McLenaghan and Spindel [11] had previously expressed the
most general commuting operator in terms of Killing-Yano tensors. 
We have here shown that, in the general case, CKY tensors of arbitrary
degree give operators that $R$-commute with the Dirac operator. It would
be interesting to know whether, in the general case, all $R$-commuting
operators are obtained in this way from CKY tensors.

Conformal Killing-Yano tensors also give rise to symmetry operators
for Maxwell's equations. In four dimensions Kalnins, McLenaghan and Williams
[12] have obtained the most general second-order symmetry operator for the
source-free Maxwell system. Their operator contains a term involving a
4-index Killing spinor. Such a term can be constructed from a CKY tensor.
It is not clear how in higher dimensions one might use a CKY tensor to
construct a symmetry operator for Maxwell's equations. Preliminary work 
suggests that one can obtain a symmetry operator from a CKY middle-form
for the middle-form generalisation of Maxwell's equations. This is to
be contrasted with the generality of the result for the Dirac operator.  

Since Maxwell's equations formally resemble the massless Dirac equations
it is perhaps worth analysing why we cannot obtain a symmetry operator
for the Maxwell system as we did in the Dirac case. We may write the
Hodge de Rham operator $\cal D$ as a `Dirac operator'[10],
$${\cal D} \equiv d-d^*	= e^a\nabla_{X_a}\,,$$
where now the Clifford action on a $p$-form is defined by
$$e^a\phi = e^a\wedge\phi +X^a\hook\phi\,.$$
We may then define $L_\omega$ to act on differential forms. (It will not
however map a $p$-form to a form of homogeneous degree.) However, we do not
have a symmetry operator since the curvature terms no longer cancel
when the Dirac operator is replaced with the Hodge de Rham operator. 
In fact one has only to add a $U(1)$ charge to the spinor connection
to see how the extra curvature terms prevent $L_\omega$ from being a
symmetry operator in that case.

\vskip 0.3true cm
{\bf Acknowledgments.} We thank Jonathan Kress for helpful discussions
on symmetry operators for Maxwell's equations.
This work was supported by the Newcastle University Research Management
Committee. 

\vskip 0.3true cm

{\bf References}
\item{[1]}
Kentaro Yano.
 Some remarks on tensor fields and curvature.
 {\it Annals of Mathematics}, {\bf 55}(2):328--347, 1952.

\item{[2]}
Shun ichi Tachibana.
 On conformal {K}illing tensor in a {R}iemannian space.
 {\it T{\^o}hoku Mathematical Journal}, {\bf 21}:56--64, 1969.

\item{[3]} 
Toyoko Kashiwada. 
 Natural Science Report, Ochanomizu University, {\bf 19}(2):67--74, 1968.

\item{[4]}
Shun ichi Tachibana and Toyoko Kashiwada.
 On the integrability of {K}illing-{Y}ano's equation.
 {\it Journal of the Mathematical Society of Japan}, {\bf 21}(2):259--265,
  1969.

\item{[5]}
G~S Hall.
 Killing-{Y}ano tensors in general relativity.
 {\it International Journal of Theoretical Physics}, {\bf 26}(1):71--81,
  1987.

\item{[6]} 
N Kamran and R G McLenaghan. 
 Symmetry operators for the neutrino and Dirac fields on curved space-time.
 {\it Physical Review D}, {\bf 30}(2):357--362, 1984.

\item{[7]} 
B Carter and R G McLenaghan.
 Generalised total angular momentum operator for the Dirac equation in curved
space-time. 
  {\it Physical Review D}, {\bf 19}(4):1093--1097, 1979.

\item{[8]} 
G F Torres del Castillo. 
 Killing spinors and massless spinor fields. 
  {\it Proceedings of the Royal Society of London A}, {\bf 400}:119--126, 1985.

\item{[9]} I M Benn, Philip Charlton and Jonathan Kress. 
 `Debye Potentials for Maxwell and Dirac Fields From a Generalisation of the
Killing-Yano Equations'. 
  Preprint, 1996 (gr-qc/9610037).

\item{[10]}
I~M Benn and R~W Tucker.
 {\it An Introduction to Spinors and Geometry with Applications in
  Physics}.
 IOP Publishing Ltd, Bristol, 1987.

\item{[11]}
R~G McLenaghan and Ph~Spindel.
 Quantum numbers for {D}irac spinor fields on a curved space-time.
 {\it Physical Review D}, {\bf 20}(2):409--413, 1979.

\item{[12]} 
E G Kalnins, R G McLenaghan and G C Williams. 
 Symmetry operators for Maxwell's equations on curved space-time. 
 {\it Proceedings of the Royal Society of London A}, {\bf 439}:103--113, 1992. 

\bye